\documentclass[twocolumn,footinbib]{revtex4}

\usepackage{amsmath}
\usepackage{graphicx}
\usepackage{float}
\usepackage{esvect}

\begin{document}
\title{Mean-field-like behavior of the generalized voter-model-class kinetic Ising model}
\author{Sebastian M. Krause}
\author{Philipp B\"ottcher}
\author{Stefan Bornholdt}
\affiliation{Institut f\"ur Theoretische Physik, Universit\"at Bremen, Otto-Hahn-Allee 1, 28359 Bremen}
\begin{abstract}
We analyze a kinetic Ising model with suppressed bulk noise which is a prominent representative of the generalized voter model phase transition. On the one hand we discuss the model in the context of social systems, and opinion formation in the presence of a tunable social temperature. On the other hand we characterize the abrupt phase transition. The system shows non-equilibrium dynamics in the presence of absorbing states. We slightly change the system to get a stationary state model variant exhibiting the same kind of phase transition. Using a Fokker-Planck description and comparing to mean field calculations, we investigate the phase transition, finite size effects and the effect of the absorbing states resulting in a dynamic slowing down. 
\end{abstract}


\maketitle

\section{Introduction}

Opinion formation models sparked considerable interest in the physics community, partly due to its close relationship to spin models (see \cite{castellano2009} for an overview). Simple mathematical rules for the outcome of discussions among agents determine the dynamics of a system of agents, as for example agreement including the time needed for consensus or disagreement, as well as the spatial spreading of opinions in terms of coarsening or segregation. Three interesting lines of research in this field are: the voter model, the universality class of the generalized voter model, and the Sznajd model. 

The Voter Model (VM) \cite{clifford1973} on regular lattices describes agents with two possible opinions, denoted as spin values $\pm 1$. The agents are randomly chosen to adopt the opinion of one of their nearest neighbors. This parameter free $Z_2$ symmetric model includes absorbing states with total agreement and thus we have a non-equilibrium system. The VM turned out to be one of the rare analytically solvable non-equilibrium models (see \cite{castellano2009} and references therein). Additional interesting properties of the model are the lack of surface tension and a slow domain growth with diffusively roughened interfaces in two dimensions \cite{dorn01}. 

The universality class of the generalized voter model (GVM) \cite{dorn01} is characteristic for systems with parameterized interactions which show a special non-equilibrium phase transition between order (consensus) and disorder for certain parameter variations. These systems may include the VM at the critical point, and exhibit an abrupt phase transition (with a jump in the order parameter at the critical parameters), however, show critical divergences with specific critical exponents for susceptibility ($\gamma=1$) and correlation length ($\nu=1/2$) \cite{dorn01}. Besides the directed percolation phase transition the GVM phase transition is  a central universality class of non-equilibrium phase transitions \cite{odor04,hin10}. Many new models showing a generalized voter transition were defined and investigated, for example using backward Fokker-Planck equations, mean field calculations, and Langevin description \cite{cast09,ham05,vazquez2008}.

The Sznajd model finally puts more emphasis on the social interpretation, where persuasiveness increases with the number of proponents \cite{sznajd2000} (see \cite{castellano2009} for an overview of dynamics and model variants). A similar effect of winning local majority is incorporated in Majority Rule models \cite{galam2002,castellano2009}. A model with agents following local as well as global majorities is used in the context of stock markets \cite{bornholdt2001,krause2011}.

We here analyze a kinetic Ising model with suppressed bulk noise \cite{oliv93}, which we believe to be an interesting paradigmatic case. On the one hand as argued in \cite{dorn01} it shows generalized voter like behavior, and it includes the voter model at its critical point. On the other hand we will discuss an effect on the level of single agents comparable to the Sznajd model, in the presence of a social temperature. 

The model is defined in section \ref{sec:model}. Its dynamics is described briefly with special emphasis on the role of the model parameter as social temperature. Finally, the presence of absorbing states is discussed. In section \ref{sec:transition} we slightly modify the model to avoid absorbing states. With this stationary state model variant we can describe the phase transition with standard methods\cite{dickman02}. We find an abrupt jump of the order parameter, but the fourth order cumulant and critical divergences at the phase transition emphasize the continuous type of the phase transition. The stationary state model variant shares the same critical exponents as found with dynamic non-equilibrium methods \cite{dorn01}. For the stationary state model variant we finally establish a Fokker-Planck description in section \ref{sec:diffusion}. We compare numerical results with analytical mean field results and thereby find a simple explanation for the behavior at the phase transition. This behavior is also found for a small world variant of the model, being closer to real social systems. With the Fokker-Planck description we also capture finite size effects. We find a divergence related to the absorbing states even in the stationary state model variant. This results in a slowing down of the dynamics. In section \ref{sec:summary} we summarize our results and give a brief outlook. 

\section{\label{sec:model}Model description and social temperature}

We investigate a kinetic Ising model \cite{oliv93} which consists of $N=L^2$ agents on a two dimensional torus with opinions $s_i=\pm 1$. Every spin adapts in random sequential update to his four nearest neighbors, and thus its state in the context of its neighborhood is adequately described by $(s_i,u_i)$ with the number of agreeing neighbors 
\begin{align}
u_i=\sum_{j \in {\rm nn}(i)}\delta_{s_i,s_j}
\end{align}
with nn$(i)$ denoting the nearest neighbors of agent $i$. Because we want to model equivalent opinions, the probabilities for spin flips $(s_i,u_i) \rightarrow (-s_i,4-u_i)$ are spin independent: $p_{(u_i) \to (4-u_i)}$. With the convention $p_{(u)\to (4-u)}+p_{(4-u)\to (u)}=1$, as in a heat bath Monte Carlo simulation of the Ising model, we immediately get $p_{(2)\to (2)}=1/2$, leaving only two independent parameters $p_{(4)\to (0)}$ (isolation, bulk noise) and $p_{(3)\to (1)}$ (join minority, interfacial noise). 

Heat bath flipping probabilities for the Ising model (with a coupling constant of one) in our notation read as $p_{(u) \rightarrow (4-u)}=(1+\exp(-\beta (8-4u))^{-1}$ with the inverse temperature $\beta$. This relates the two independent parameters to each other and leads to an equilibrium system with defined temperature obeying detailed balance for single spin flips \cite{oliv93} (by construction of the heat bath algorithm). In kinetic Ising models this relation is not fulfilled. Such models can be understood as being coupled to two heat baths of different temperatures \cite{oliv93} and thus they experience a flux of heat, leading to a first sign of non-equilibrium behavior, which will be supplemented by a more drastic non-equilibrium property later on. 

\begin{figure}[htb]
\begin{center}
	\includegraphics[width=0.7\columnwidth]{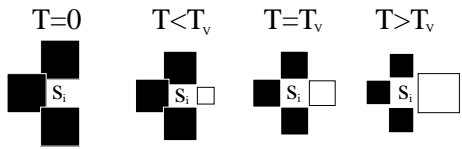}
	\caption{Illustration of the persuasiveness an agent $s_i$ receives from its neighbors for different temperatures $T$. For low temperatures, majorities convince stronger. For high temperatures, the agent follows any present opinion in a panic-like mood.}
		\label{fig:persuasive}
\end{center}
\end{figure}
We here introduce a system without voluntary isolation of agents which means the lack of bulk-noise or in other words zero temperature of bulk noise. This leads to the transition probabilities
\begin{eqnarray}
&&p_{(4) \rightarrow (0)}  =  0  \\
&&P_{(0) \rightarrow (4)}  =  1  \nonumber \\
&&\text{(no isolation / no bulk-noise)}	\nonumber \\
&&p_{(3) \rightarrow (1)}  =   \frac{1}{1+\exp(4 \beta)} \\
&&p_{(1) \rightarrow (3)}  =  1-p_{(3) \rightarrow (1)} 	\nonumber \\
&&\text{(interfacial-noise).} 	\nonumber
\label{interface}
\end{eqnarray} 
This single parameter model for any $\beta$ shares the property of the voter model, where agents only adopt opinions which actually exist in their neighborhood. This property is reasonable since total isolation seems to be a quite rare event in social systems. In the voter model every neighbor influences an agent with the same persuasiveness, since one of his neighbors is randomly chosen for his update. Thus we get $p_{(u) \rightarrow (4-u)}=(4-u)/4$, which is a special case in our model defined by $p_{(3) \rightarrow (1)}=1/(1+\exp(4/T_{\rm V} ))=1/4$ with the special temperature $T_{\rm V} = 4/\ln(3) \approx 3.641$. For smaller temperatures the model shows an increased persuasiveness of groups of agents, $p_{(1) \rightarrow (3)} > 3 p_{(3) \rightarrow (1)}$, as illustrated in Fig.\ \ref{fig:persuasive}. This motivates the name Group-Voter Model (GRVM). This effect, which motivated models like the Sznajd model and majority models \cite{castellano2009}, is incorporated in our model in a continuously tunable way. For higher temperatures $T>T_{\rm V}$, local majorities have a suppressed persuasive power. Agents adopt any opinion in their neighborhood without trusting majorities which can be seen as a panic-like behavior. 

\begin{figure}[htb]
\begin{center}
	\includegraphics[width=1.0\columnwidth]{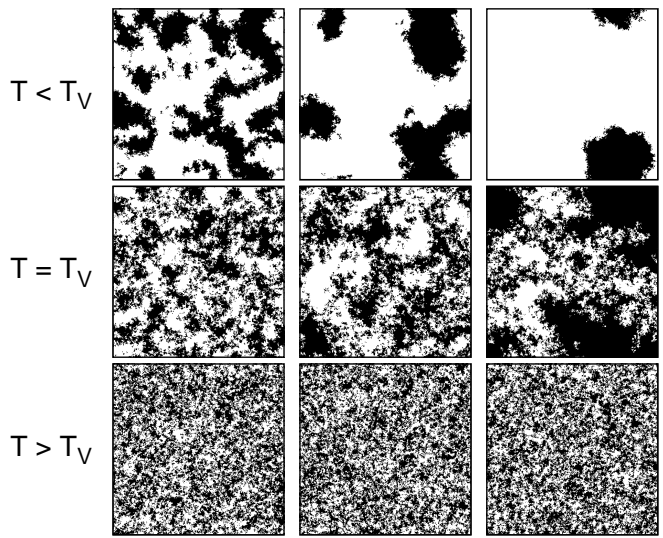}
	\caption{Lattice states (L = 500) for $T=0.9\,T_{\rm V}$ (top), $T=T_{\rm V}$ (middle) and $T=1.1\,T_{\rm V}$ (bottom) at increasing times $t=1000$ (left), $t=10000$ (middle), and $t=50000$ (right). The changing ordering dynamics indicates a phase transition at $T_{\rm V}$, where the dynamics is identical to the voter model.}
		\label{evol}
\end{center}
\end{figure}
In Fig.\ \ref{evol}, the coarsening dynamics starting from random initial conditions can be seen for the different temperature regimes and a system of $N=500^2$ agents. For low temperatures (top line), where local majorities are strongly preferred and thus single agents follow group opinions, we see ordering dynamics leading to an ordered phase. So the group-following tendency has an ordering effect compared to the pure voter model. At $T=T_{\rm V}$ we get the voter model with slow cluster growth and roughened surfaces. At this temperature we find a phase transition from order to disorder (see below). At this point the temporal decline of the density of surfaces changes from power law behavior through logarithmic decline to saturation \cite{dorn01}. For $T>T_V$ we see a disordered noisy system, so the lack of trust of single agents in local majorities has a destabilizing effect. 

Summarizing, an increasing temperature leads to a decreasing group effect (Fig.\ \ref{fig:persuasive}), and simultaneously it leads to an increased dynamical temperature (Fig.\ \ref{evol}). This can be interpreted as follows: In a faster changing environment, any opinion provided by any neighbor seems interesting, single neighbors might be right by chance. In the limit of infinite temperature, opinions which are present in the neighborhood lead to adaption with probability $1/2$ regardless of minorities or majorities. On the other hand, decreasing trust in local majorities indeed accelerates the dynamics, so the described effect is self consistent. The complementary micro- and macro-effects of the parameter $T$ ranging from group following and ordered states to panic reactions and disordered states allow us to interpret it as a social temperature. 

For $T>T_V$ the system remains in a state in which the magnetization per site 
\begin{align}
	m &= \frac{1}{N} \displaystyle\sum_{j}^{} s_j
\end{align}
oscillates around zero. Note that the amplitude of the oscillations is smaller for higher temperatures and higher system sizes. 
\begin{figure}[htb]
\begin{center}
		\includegraphics[width=1.0\columnwidth]{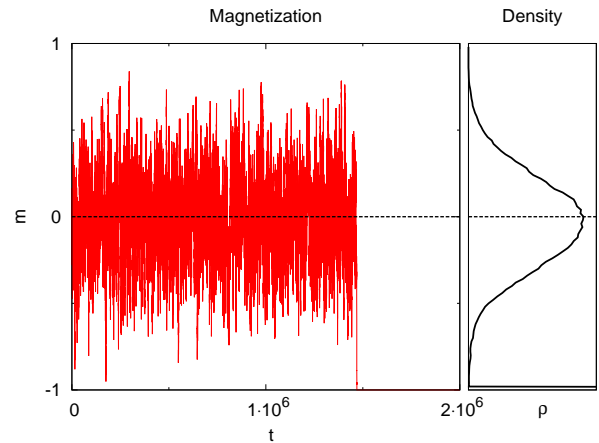}
		\caption{(Color online) Left: Magnetization over time $t$ for $T = 1.02\,T_V$ and $L = 75$. Right: Magnetization density. Absorbing states can be observed.}
		\label{isola}
\end{center}
\end{figure}
In Fig.\ \ref{isola} we finally see the presence of absorbing states for a system with $L=75$ and $T = 1.02 \cdot T_V$. The system is in the disordered state and oscillates around zero magnetization for more than one million sweeps, and the according density of magnetization 
\begin{align}
	\rho(m') &= \frac{1}{c}\displaystyle\sum_{t}^{} \delta_{m',m(t)}
\end{align} 
with the normalization constant $c$ and the Kronecker $\delta_{i,j}$ suggests that the system reached a stationary state. 
Nevertheless the system finally gets trapped in an absorbing state. These absorbing states exist due to the absence of bulk noise. In finite systems they are reached for any temperature and system size. Thus the system starting from random initial conditions performs a transient nonequilibrium dynamics. 

The description of non-equilibrium systems is somewhat complicated. There are different approaches to deal with the properties associated with non-equilibrium systems. One possible way is to investigate the phase ordering dynamics and recover dynamic critical exponents \cite{dorn01}.

We here use a different procedure. We slightly change the dynamics to prevent the system from reaching the absorbing state. The resulting stationary state model variant we will describe and investigate in the next section. 

\section{\label{sec:transition}Stationary state model variant and the phase transition}

\begin{figure}[htb]
\begin{center}
\includegraphics[width=1.0\columnwidth]{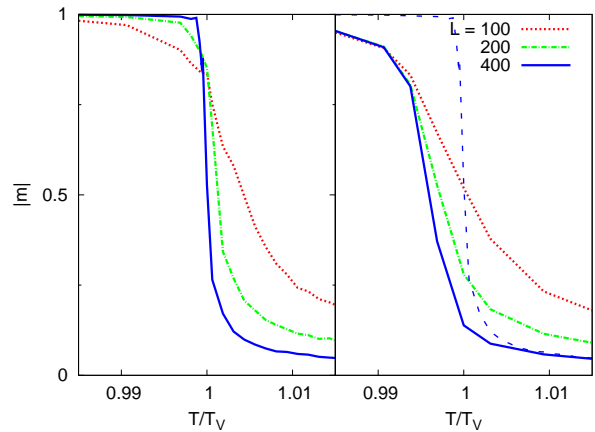}
\caption{(Color online) Comparison of the stationary state model variants with the daemon rule (left) and with bulk noise ($\alpha = 10^{-4}$, right) with the dashed line (daemon rule for $L=400$) for comparison. The daemon rule reproduces the jump of the control parameter in the thermodynamic limit.}
\label{devil}
\end{center}
\end{figure}
Similar to \cite{munoz98}, we use a minimal modification of the original model to get an estimation of the quasi-static properties. We let a little daemon keep the last remaining spin from flipping which prevents the system from reaching the absorbing states. The modified model reaches a stationary state which is used for the evaluation of the statistical quantities. As shown by Dickman and Vidigal \cite{dickman02}, such methods are suitable for studying the universal behavior of transient dynamics that otherwise would decay into absorbing states. As we still have a system coupled to two heatbaths with different temperatures, our system exhibits a non-equilibrium stationary state (NESS).

We show that the daemon rule indeed is minimal by comparing it to a different procedure. In the case of a non zero isolation probability $p_{4 \rightarrow 0} = \alpha p_{3 \rightarrow 1}$ with little $\alpha$ a little bit of bulk-noise is introduced. The magnetization for different $L$ is shown on the right hand side of Fig.\ \ref{devil}. The phase transition of the system with tiny bulk noise is Ising-like. The smooth transition can also be observed for bigger $L$, although $\alpha = 10^{-4}$ is already considerably small. This suggests that the limit $\alpha \rightarrow 0$ in addition to $L \rightarrow \infty$ must be taken to model the phase transition of the GRVM. Whereas the daemon rule produces a phase transition which displays the abrupt change in magnetization $m$. This is shown on the left hand side of Fig.\ \ref{devil}. Note that the impact of the daemon rule decreases for bigger $L$, whereas the impact of the method with bulk noise remains constant.

\begin{figure}[htb]
\begin{center}
		\includegraphics[width=1.0\columnwidth]{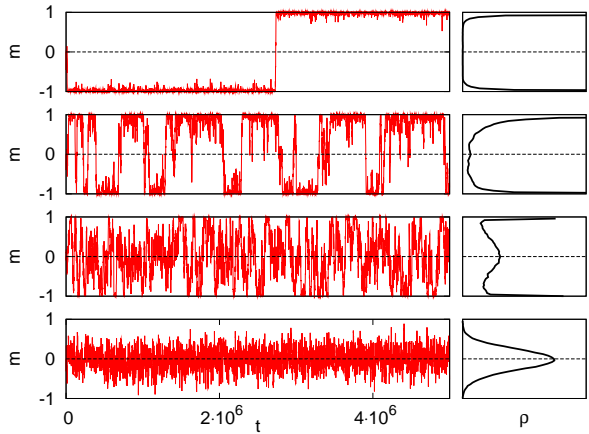}
		\caption{(Color online) Comparison of magnetization $m$ over time $t$ (left) and its density $\rho$ (right) for different temperatures $T$ for $L=100$. From top to bottom: $T=0.9846\,T_V$, $T=1.0002\,T_V$, $T=1.0046\,T_V$ and $T=1.6717\,T_V$. The time spend near the absorbing states is long for $T \approx T_V$, giving rise to a slowdown of the simulation.}
		\label{slowdown}
\end{center}
\end{figure}
Although this method eliminates the existence of absorbing states and produces a system with suitable stationary state, the absence of bulk-noise gives rise to another problem. If the system is in a state where the daemon is needed, it is likely that the system will remain in this state for an extended amount of time. 
Time series of $m$ and the corresponding densities for four different $T$ with 5 million sweeps and $L=100$ are shown in Fig.\ \ref{slowdown}. The absorbing states dominate the time series of $m$ for $T<T_V$. For $T\approx T_V$ the system spends a finite time in states where $|m|$ is not close to 1, but states with $|m|$ close to 1 are still the most frequent states and the system spends a typical time of approximatively 1 million sweeps near one of the absorbing states. If we consider the plots corresponding to $T=1.0046 \cdot T_V$, we see that although a clear maximum in the magnetization's density $\rho$ can be found at $m=0$, the system spends a considerable time in a state in which $|m|$ is close to one. This effect is a sign of the absorbing state around $|m|=1$ due to vanishing diffusion, as will be discussed in the next section. The effect increases with system size and thus facilitates finite size scaling. It must be taken into account additionally to the critical slow down and it dramatically enhances the number of sweeps needed for good statistics. 

\begin{figure}[htb]
\begin{center}
		\includegraphics[width=1.0\columnwidth]{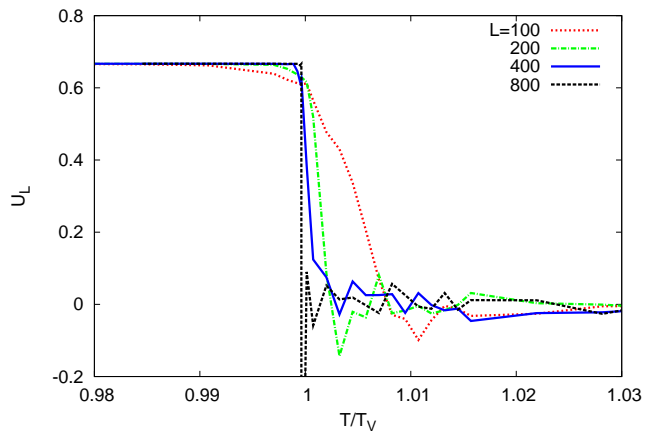}
		\caption{(Color online) The fourth order cumulant $U_L$ over the temperature for different system sizes $L^2$. The typical behavior for a continuous phase transition can be observed.}
		\label{fourthcumu}
\end{center}
\end{figure}
To examine the critical behavior at the phase transition we first observe the fourth order cumulant $U_L$. We evaluate $U_L$ with the time series of the magnetization using
\begin{align}
	U_L &= 1-\frac{\langle m^4 \rangle}{3 \langle m^2\rangle^2}.
\end{align}
$U_L$ can be used to determine the order of transition \cite{binder}: A first order transition would have a $U_L$ that displays a clear minimum. This minimum gets closer to the system's critical parameter as $L$ gets bigger.
A second order transition has a $U_L$ that starts at $\frac{2}{3}$ and falls monotonously to zero as $T$ grows. The $U_L$ intersects at the critical parameter of the system.

The calculated fourth order cumulant (Fig.\ \ref{fourthcumu}) is $\frac{2}{3}$ for $T<T_V$ and falls to zero for $T>T_V$. This behavior is in agreement with a second order phase transition. The system has therefore a discontinuous phase transition with critical behavior as typical for continuous phase transitions. 

\begin{figure}[htb]
\begin{center}
		\includegraphics[width=1.0\columnwidth]{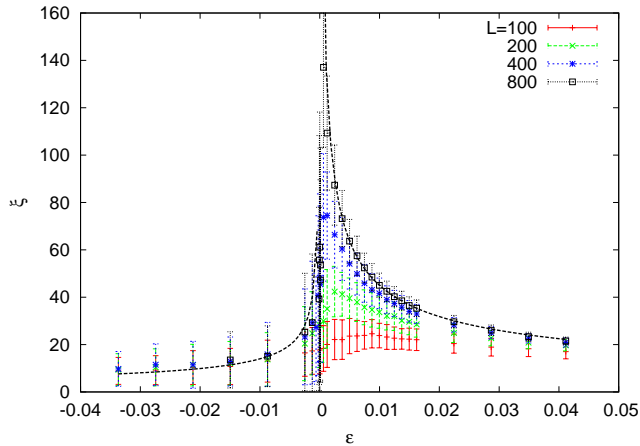}
		\caption{(Color online) Correlation length $\xi$ for different system sizes $L^2$ over reduced temperature $\varepsilon$. Dashed line: Plot of $\xi \propto \varepsilon^{-1/2}$ to compare the data to the expected critical exponent.}
		\label{correl}
\end{center}
\end{figure}
Due to the phase transition with continuous properties a critical behavior with scaling laws is expected. We  evaluate the susceptibility and the correlation length to determine the scaling exponents $\gamma$ and $\nu$. 
The correlation length $\xi$ can be measured by evaluating the structure factor \cite{newmann}, which is often used in solid states physics. The structure factor is obtained using $S(\vec{k}) =  | \tilde{s}'(\vec{k})|^2$ where $\tilde{s}'= F\{s'\}$ is the Fourier transformed grid $s' = s-m$. Note that this definition ensures that states with large magnetization lead to small values of $\xi$.  We take the circular average over $k$ and get $S(k)$. The structure factor should go to zero as the magnitude of $\vec{k}$ approaches zero or infinity and thus $S(k)$ displays a peak at the $k'$ corresponding to the average domain size. $\xi = 2\pi/k'$ was calculated by using 
	\begin{align}
  	  k' &=  \frac{\int_0^{\infty} k \cdot S(k) dk}{\int_0^{\infty} S(k) dk}.
	\end{align}

The simulations were performed with 5 million sweeps. After the system reached its stationary state, $\xi$ was calculated every $10,000$ sweeps to ensure uncorrelated results. An average of the domain size $\xi$ and the error given by the standard deviation was therefore possible to acquire. The error near the critical point is relatively large. One could assume that the errors would be smaller if $\xi$ was calculated more often, for example every $100$ sweeps. This is not the case, since the discussed slowdown has to be considered. As shown in
Fig.\ \ref{slowdown}, the time required to go through a sufficient number of states is large for $T \approx T_V$. So only larger time series can improve the statistics.

If we plot the correlation length over the reduced temperature ($\varepsilon=\frac{T-T_V}{T_V}$), we should observe the typical critical behavior $\xi \propto |\varepsilon|^{-\nu} $. The exponents obey the relation
\begin{align}
	\gamma &= d\cdot \nu - 2 \beta .
\end{align} 
Since there is a jump in the order parameter ($\beta=0$) and the dimension is $d=2$, the remaining exponents obey $\frac{\gamma}{\nu} = 2$.
We expect $\nu = \frac{1}{2}$ for the universality class of the GVM \cite{dorn01}. As shown in Fig.\ \ref{correl} an exponent of $\nu = \frac{1}{2} $ fits the obtained data nicely.

\begin{figure}[htb]
\begin{center}
		\includegraphics[width=1.0\columnwidth]{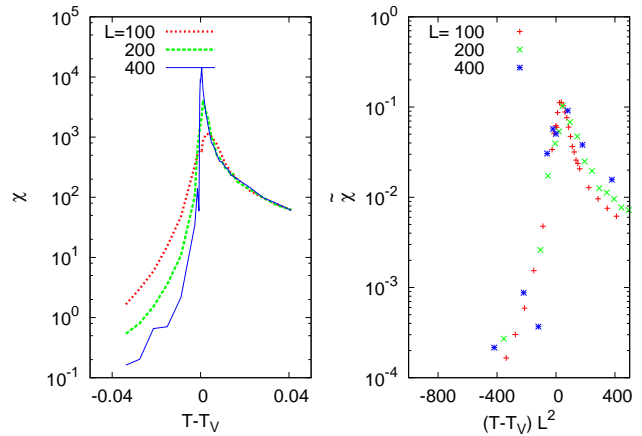}
		\caption{(Color online) Left: susceptibility $\chi$ as a function of $T-T_{\rm V}$. Right: reduced susceptibility $\tilde{\chi}$ as a function of $(T-T_{\rm V})L^2=\tilde{\varepsilon}\,T_{\rm V}$ with $\frac{\gamma}{\nu} = 2$ and $T_C = 0.9997\,T_V$}
		\label{sus}
\end{center}
\end{figure}
To evaluate the remaining critical exponent $\gamma$ the susceptibility $\chi_L$ is needed. We calculate $\chi_L$ as proposed in \cite{oliv92} by using the time series of $m$, 
\begin{align}
	\chi_{L} &= N (\langle m^2 \rangle - \langle |m|\rangle^2) \propto |\varepsilon|^{-\gamma}.
\end{align} 
The exponent $\gamma$ can be found by introducing the reduced susceptibility given by
\begin{align}
	\tilde{\chi}(\tilde{\varepsilon}) &= \tilde{\chi}(\varepsilon L^{\frac{1}{\nu}}) = \chi_{L}(\varepsilon) L^{-\frac{\gamma}{\nu}}. 
\end{align} 
The reduced susceptibility $\tilde{\chi}$ plotted against $\tilde{\varepsilon}=\varepsilon L^{\frac{1}{\nu}}$ for different $L$ should collapse for $T < T_V$ given the correct $\gamma$, $\nu$ and critical temperature $T_C$. This was realized for different $L$ (Fig.\ \ref{sus}). The chosen parameters to obtain Fig.\ \ref{sus} were $\frac{\gamma}{\nu} = 2$ and $T_C = 0.9997\,T_V$. 

\section{\label{sec:diffusion}Fokker-Planck equation and mean field description}

As a starting point for the Fokker-Planck description of the dynamics, we use transition probabilities for the magnetization $p^\pm(m)\equiv p_{m \pm \Delta m,m}$ for single time steps $\Delta t=1/N$ with $\Delta m=2/N$, which describes a single possibly flipping spin. With these two switching probabilities at hand, the Fokker-Planck equation can be deduced in a very simple way as described in the following. We analyze the time evolution of the probability density of the magnetization $P(m,t)$ by performing a Taylor expansion ($\Delta P=P(m,t+\Delta t)-P(m,t)$)
\begin{eqnarray}
\frac{\Delta P}{\Delta t} &=&  \frac{p^+(m-\Delta m)}{\Delta t}P(m-\Delta m,t) \nonumber \\
& & + \frac{p^-(m+\Delta m)}{\Delta t}P(m+\Delta m,t) \nonumber\\
& & - \left(\frac{p^+(m)}{\Delta t}+\frac{p^-(m)}{\Delta t}\right)P(m,t) \nonumber\\
&\approx&  -\frac{\partial}{\partial m} \left(\frac{\Delta m}{\Delta t}(p^+(m)-p^-(m))P(m,t)\right) \\
& & + \frac{1}{2}\frac{\partial^2}{\partial m^2} \left(\frac{{\Delta m}^2}{\Delta t}(p^+(m)+p^-(m))P(m,t)\right). \nonumber
\end{eqnarray}
This leads to the Fokker-Planck equation
\begin{align}
\frac{\partial P}{\partial t} & = -\frac{\partial}{\partial m} \left(a_1(m)P(m,t)\right) + \frac{1}{2}\frac{\partial^2}{\partial m^2} \left(a_2(m)P(m,t)\right)\\
a_1 &= \frac{\Delta m}{\Delta t}(p^+(m)-p^-(m)) \\
a_2 &= \frac{{\Delta m}^2}{\Delta t}(p^+(m)+p^-(m))
\end{align}
with the drift term $a_1(m)$ and the diffusion term $a_2(m)$. Using a potential for the drift term $a_1(m)=-\frac{\rm d}{{\rm d} m}V(m)$ the phase transition can be understood. Additionally with the diffusion term $a_2(m)$ and the stationary solution
\begin{align}
P_{\rm s}(m)=\frac{c}{a_2(m)}\exp\left(2\int_{m_0}^m \frac{a_1(m')}{a_2(m')} {\rm d}m'\right)
\end{align}
with the normalization constant $c$, finite size effects may be investigated. 

To calculate the properties of interest, the transition probabilities $p^{\pm}(m)$ are needed. For simulations on the grid, they are calculated using the time series $m(t)$ (with single step resolution $\Delta t=1/N$). With $\Delta m(t)=m(t+\Delta t)-m(t)$ we get 
\begin{align}
p^{\pm}(m)=\frac{\sum_t \delta_{\Delta m(t),\pm \Delta m}\delta_{m(t),m}}{\sum_t \delta_{m(t),m}}.
\label{eq:measure}
\end{align}

We want to compare our results to mean field calculations, which we calculate using the flipping probabilities $p_{(u) \rightarrow (4-u)}$ and
\begin{align}
p^\pm(m)=\sum_{u=0}^4 \frac{N_{(\mp,u)}(m)}{N}p_{(u) \to (4-u)}
\end{align}
with $N_{(s,u)}$ being the total number of spins with  the state $(s,u)$. Assuming $N_{(s,u)}$ to have a strict dependence on the magnetization is reasonable for the mean field case, where we get (including the unlikely case of twice choosing the same neighbor)
\begin{align}
N_{(s,u)}^{\rm MF}(m) &=\frac{N_{(s)}}{N}\cdot \left(\frac{N_{(s)}}{N}\right)^u \cdot \left(\frac{N_{(-s)}}{N}\right)^{4-u}\\
&= (\frac{m+s}{2})^{u+1}\cdot (\frac{m-s}{2})^{4-u}
\end{align}
and thus finally calculate 
\begin{align}
a_1^{\rm MF}(m)&=-2\left(p_{(3) \to (1)}-\frac{1}{4}\right)m(1-m^2)\\
V^{\rm MF}(m)&=\phi_{v}\left(p_{(3) \to (1)}-\frac{1}{4}\right)\left(m^2-\frac{m^4}{2}\right)+v_0\label{eq:MFV}\\
a_2^{\rm MF}(m)&=\frac{2}{N}\left((1-m^2)+\left(p_{(3) \to (1)}-\frac{1}{4}\right)m^2(1-m^2)\right)\nonumber\\
&\approx \phi_{a}\frac{2}{N}(1-m^2)\label{eq:MFa2}\\
P_{\rm s}^{\rm MF}(m)&\approx \frac{c}{1-m^2}\exp\left(-\phi_{r}\left(p_{(3) \to (1)}-\frac{1}{4}\right)m^2N\right)\label{eq:MFrho}
\end{align}
with approximate results for $\left|p_{(3) \to (1)}-1/4 \right| \ll 1$. The parameters $\phi$ are needed for a fit procedure later on. They are $\phi_{v}=\phi_{a}=\phi_{r}=1$, $v_0$ may be chosen and $c$ is a normalization constant.

\begin{figure}[htb]
\begin{center}
		\includegraphics[width=1.0\columnwidth]{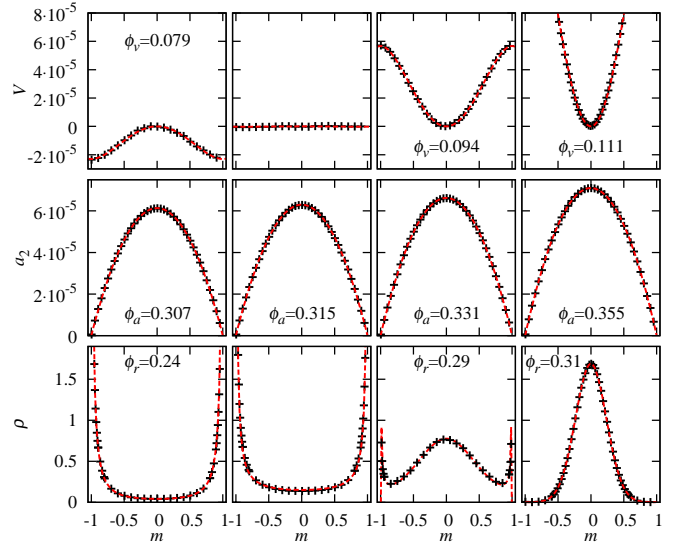}
		\caption{(Color online) Drift potential $V(m)$ (top), diffusion term $a_2(m)$ (middle) and density $\rho(m)$ (bottom) for different temperatures (from left to right $T=0.997\,T_{\rm V}$, $T=T_{\rm V}$, $T=1.006\,T_{\rm V}$, $T=1.016\,T_{\rm V}$) and $L=100$. Measured data (+) is compared to mean field results with good agreement. The fitting parameters say that diffusion and drift are suppressed compared to the mean field case. The drift potential clarifies the phase transition. The vanishing drift at $|m|=1$ leads to a divergence in the density, related to the absorbing states connected with a dynamic slowing down.}
		\label{fig:fokkerplanck}
\end{center}
\end{figure}
The symbols in Fig. \ref{fig:fokkerplanck} show results for the drift potential $V(m)$ (top), the diffusion term $a_2(m)$ (middle) and the density $\rho(m)$ (bottom) for the following temperatures around the critical temperature (from left to right): $T=0.997\,T_{\rm V}$, $T=T_{\rm V}$, $T=1.006\,T_{\rm V}$, $T=1.016\,T_{\rm V}$. The system size was chosen as $N=L^2=100^2$ and for the lower temperatures time series of $6 \cdot 10^8$ sweeps were performed, for the higher temperatures $2 \cdot 10^8$ sweeps sufficed. So we used altogether more than $10^{12}$ steps in eq. (\ref{eq:measure}). The dashed lines in the figure are fits of the mean field functions (\ref{eq:MFV}), (\ref{eq:MFa2}) and (\ref{eq:MFrho}) to the measured data and the fit parameters $\phi_{v}$, $\phi_{a}$ and $\phi_{r}$ are given (two of them are not defined in the critical case). As we can see, drift and diffusion are suppressed compared to the mean field case, which leads to an effect to the density, which might be seen as a temperature stretching procedure away from the critical temperature $T_{\rm V}$. Apart from that, the system shows perfect mean field like behavior. With this knowledge we are able to discuss the phase transition including the critical exponents and the finite size properties of the system including the role of the absorbing states. 

To discuss the phase transition, we have to consider the case $N \to \infty$. From calculations performed for $N=20^2$ and $N=50^2$ we know, that the system behaves mean field like in all these cases, with shrinking parameters $\phi_{v}$, $\phi_{a}$ and $\phi_{r}$ for increasing $N$. The drift correction shrinks faster than the diffusion correction, however, diffusion is additionally proportional to $1/N$ and thus the influence of fluctuations shrinks with increasing system size. In the thermodynamic limit we thus have to consider the minima of the potential $V(m)$ as shown in the top line of Fig.\ \ref{fig:fokkerplanck} (see also \cite{ham05,cast09,vazquez2008}). The position of the minima performs a jump from $|m|=1$ to $m=0$ at $T=T_{\rm V}$ and thus we have an abrupt phase transition with exponent $\beta=0$. At the temperature $T_{\rm V}$ the potential vanishes which corresponds to the vanishing drift in the voter model. Adding a term $-h m$ to the potential in eq. (\ref{eq:MFV}), we can calculate the susceptibility $\chi=\frac{\partial m}{\partial h}$ by calculating the minimum $m(h)$ and thus prove the mean field exponent $\gamma=1$, because near the phase transition we get $\chi\propto \left(p_{(3) \to (1)}-\frac{1}{4}\right)^{-1}\propto (T-T_{\rm C})^{-1}$ (strictly speaking the magnetization $m=\pm 1$ in the ordered phase may only be influenced by a field $h$ of the opposite sign). 

Finite size effects are especially visible in the densities as can be seen in the bottom line of Fig.\ \ref{fig:fokkerplanck} and in eq. (\ref{eq:MFrho}), respectively. Integrating the density, the term $(1-m^2)^{-1}$ resulting from the vanishing diffusion at $m=\pm 1$ (with simultaneously vanishing drift) leads to a divergence (compare \cite{dickman02,munoz98}). In the simulations the daemon keeps last spin from flipping and thus the integration limits are reduced to $|m|=1-2/N$. Nevertheless, the absorbing states can be seen around $|m|=1$, as for increasing system size $N$, the system with a temperature near the phase transition (and even in the disordered phase, as can be seen for $T=1.006\,T_{\rm V}$ in Fig.\ \ref{fig:fokkerplanck}) will spend an increasing amount of time near the absorbing state. So we again get a system which is trapped by the absorbing state if the system once moves to this state. This causes a slowing down in simulations, which is even more influential than the critical slow down in equilibrium systems and results in a poor convergence. 

Finally we want to discuss a small world variant of the system, where some of the nearest-neighbor grid connections are rewired. It is quite meaningful to cover this case in the social context, since social networks are known to have small world properties. Because the small world network is somewhere in between the grid (large diameter) and the mean field case (diameter of one), and the results on the grid are qualitatively mean field like, the results for the small world variant should be mean field like, as well. 

\begin{figure}[htb]
\begin{center}
		\includegraphics[width=1.0\columnwidth]{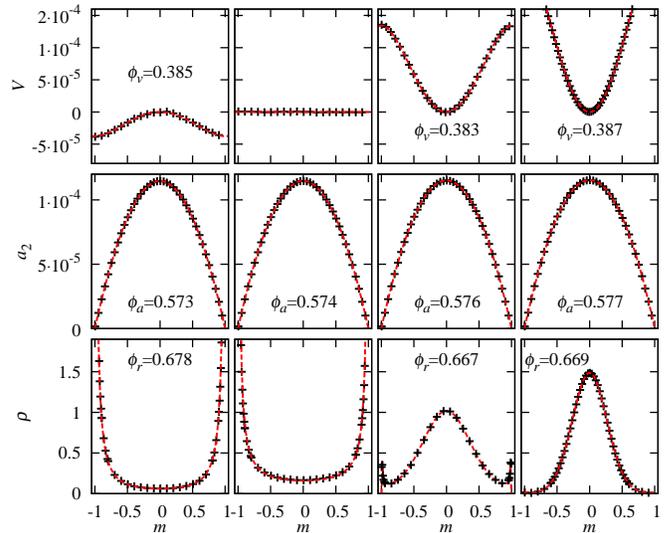}
		\caption{(Color online) As in Fig.\ \ref{fig:fokkerplanck} for a small world variant with 5\% rewired links and temperatures (from left to right) $T=0.998\,T_{\rm V}$, $T=T_{\rm V}$, $T=1.003\,T_{\rm V}$, $T=1.006\,T_{\rm V}$. The shortcuts lead to increased drift and diffusion.}
		\label{fig:smallworld}
\end{center}
\end{figure}
We start from the grid and use directed links, which are directed from influential agents to the influenced ones. So an agent takes into account four in-links to calculate the switching probabilities, or more precisely the opinions of the agents on the other end of the in-links. We rewire 5\% of the in-links by randomizing the starting points of the links and keeping the end points, which keeps all in-degrees to be four and only changes the out-degree of several agents (since single agents might have an out degree of zero, we have to change the daemon rule to act on more than one last spin). Fig.\ \ref{fig:smallworld} shows results for the small world network as in Fig.\ \ref{fig:fokkerplanck}, here for the temperatures (from left to right) $T=0.998\,T_{\rm V}$, $T=T_{\rm V}$, $T=1.003\,T_{\rm V}$ and $T=1.006\,T_{\rm V}$. We find the same mean field like behavior as for the grid and conclude, that the same kind of phase transition should appear. The fit parameters $\phi$ are more close to the mean field case $\phi_{v}=\phi_{a}=\phi_{r}=1$ and we found using calculations for $N=50^2$ that these fit parameters do not remarkably change for increasing system size, which is in accordance to slowly growing diameters in small world networks. 

\section{\label{sec:summary}Summary}

We investigate a kinetic Ising model with suppressed bulk noise, which is known as a prominent representative of the generalized voter model phase transition. For low temperatures, we emphasize the effect of enhanced persuasiveness of groups on the level of single agents. This leads to ordered states compared to the pure voter model. Similar effects are implemented in opinion formation models as for example the Sznajd model or the Majority Rule model. For high temperatures the opposite effect of lacking trust in majorities can be observed which leads to increasingly disordered states. Through the model parameter $T$ the group effect can be tuned, which leads to a changing behavior of single agents and the system as a whole as well. This effect allows us to identify the model parameter as a social temperature. So we find that this single parameter system shows interesting properties not only for its behavior at the phase transition, but also provides intuitive rules at the level of single agents. 

To describe the system using its transient quasi-static properties, we change the dynamics using a minimal rule which prevents that absorbing states are reached. We find the generalized voter model transition, as it has been found for the original system \cite{dorn01}. With the fourth order cumulant we emphasize the continuous type of the phase transition despite the jump in the order parameter.  

Finally we derive the Fokker-Planck description of the phase transition. We measure drift and diffusion using numeric time series and compare it to analytic mean field results. We find perfect mean field behavior, only with suppressed diffusion and drift. With this we understand the abrupt phase transition including the value of the critical exponent $\gamma$. We additionally find a divergence in the magnetization density due to vanishing drift and diffusion, related to the absorbing states. This leads to a dynamic slowing down. 

As a system between grid and mean field, we also check a small world variant. This system shows exactly the same mean field like behavior, but diffusion and drift are less suppressed. This system is more similar to real social systems with small world properties. 

In summary we obtain a physical understanding of the voter model class in terms of a kinetic Ising model, obeying mean field dynamics.

\end{document}